\documentclass[%
 reprint,
 amsmath,amssymb,
 aps,
]{revtex4-1}
\usepackage{graphicx}% Include figure files
\usepackage{dcolumn}% Align table columns on decimal point
\usepackage{bm}% bold math
\usepackage{graphicx}% Include figure files
\usepackage{dcolumn}% Align table columns on decimal point
\usepackage{bm}% bold math
\usepackage{bigints}
\usepackage{color}
\usepackage{amsmath,amssymb,graphicx}
\usepackage{float}
\usepackage{braket}
\usepackage{tabularx}
\usepackage{tabulary}
\usepackage{tabu}
\usepackage{booktabs}
\usepackage{textcomp}
\usepackage{graphicx}% Include figure files
\usepackage{dcolumn}% Align table columns on decimal point
\usepackage{bm}% bold math
\usepackage{graphicx}% Include figure files
\usepackage{dcolumn}% Align table columns on decimal point
\usepackage{bm}% bold math
\usepackage{bigints}
\usepackage{color}
\usepackage{amsmath,amssymb,graphicx}
\usepackage{amssymb}
\usepackage{gensymb}
\usepackage{amsfonts}
\usepackage{float}
\usepackage{braket}
\usepackage{tabularx}
\usepackage{tabulary}
\usepackage{tabu}
\usepackage{booktabs}
\usepackage{textcomp}
\usepackage{cleveref}
\usepackage{cancel}
\usepackage{braket}

\newcolumntype{C}{>{\centering\arraybackslash}X}
\begin{document}
\preprint{APS/123-QED}
\title{Analytical formulation of high-power Yb-doped double-cladding fiber laser}

\email{mpeysokhan@unm.edu}
\author{Mostafa Peysokhan$^{1,2,*}$}

\author{Esmaeil Mobini$^{1,2}$}
\author{Arash Mafi$^{1,2}$}

\affiliation{$^1$Department of Physics \& Astronomy, University of New Mexico, Albuquerque, NM 87131, USA \\
             $^2$Center for High Technology Materials, University of New Mexico, Albuquerque, NM 87106, USA} %Lines break automatically or can be forced with \\

%\affiliation{}%
%\date{\today}% It is always \today, today, but any date may be explicitly specified
\begin{abstract}
Here a detailed formalism to achieve an analytical solution of a lossy high power Yb-doped silica fiber laser is introduced. The solutions for the lossless fiber laser is initially attained in detail. Next, the solution for the lossy fiber laser is obtained based on the lossless fiber laser solution. To examine the solutions for both lossless and lossy fiber laser two sets of values are compared with the exact numerical solutions and the results are in a good agreement. Furthermore, steps and procedures for achieving the final solutions are explained clearly and precisely.
\end{abstract}
%%%%%%%%%%%%%%%%%%%%%%%%%%%%%%%%%%%%%%%%%%%%%%%%%%

\maketitle

\section{Introduction}
Over the past five decades following the first demonstration of glass fiber lasers by Snitzer~\cite{Koester:64}, fiber lasers have excelled in all performance attributes. The operating wavelengths cover a broad range, from ultraviolet to mid-infrared~\cite{Shi, moulton2009tm, nilsson2004high, zervas2014high, Pask, Allain, aghbolagh2019mid}, and high-power Yb-doped double-cladding fiber lasers (YDCFL) are one of the primary sources of high-power radiation for industrial and directed energy applications~\cite{Richardson:10, Zervas}. Fiber lasers can deliver powers on the order of a few kilowatts (kW)~\cite{Jeong:09, Yan, Xiao:12, jauregui2013high, Beier:17} and are promising candidates for coherent beam combining to achieve even higher powers~\cite{Klingebiel:07, Liu2013, Zheng:16}. 

Fiber lasers are typically designed using extensive numerical solutions and optimizations. It is often the case that a broad range of parameters over a multi-dimensional design space needs to be covered to find the optimal parameters for the best performance. The availability of an analytical solution would simplify the design problem significantly; moreover, it can provide a more intuitive design platform compared with brute-force numerical solutions. There are a few published papers in the literature that present analytical solutions to the fiber laser equations~\cite{Kelson, Kelson:99, xiao2004approximate, Yahel:06, duan2007analytical, mohammed2014approximate, mohammed2014optimization}. However, either the parasitic background absorption is absent in the formulation, or the results are not presented in a fully analytical closed form. In particular, it is quite essential to include the parasitic background absorption because it is one of the main detriments in modern high-power laser designs, and significant effort is put into reducing its contribution to the heating problem in fiber lasers~\cite{jetschke2008efficient, Suzuki:09, Jauregui:16}.   

In this paper, we present a complete analytical solution of the YDCFL considering the parasitic background absorption. We treat the parasitic background absorption as a perturbation. We first obtain an analytical solution to the pump and signal propagation problem in the absence of loss and then treat the parasitic background absorption using the first-order perturbation theory. We verify the accuracy of the solutions by comparing them with the full numerical solutions and show that our analytical treatment accurately models the full laser design problem in the  regimes of interest to high-power fiber laser design.
%%%%%%%%%%%%%%%
%%%%%%%%%%%%%%%
\section{Laser equations}
\begin{figure}[htp]
\centering
\includegraphics[width=3.3 in]{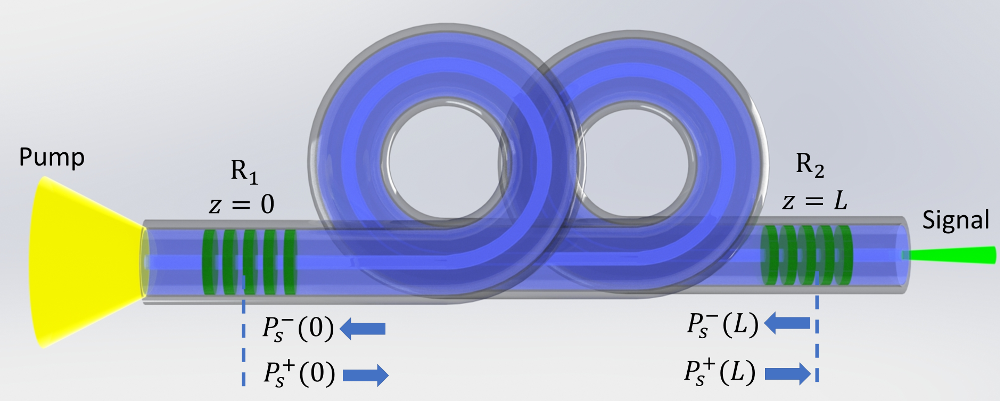}
\caption{Schematic of the laser system and propagation of the pump power and signal in the double-cladding fiber laser.
Pump power is launched at z = 0 and the output signal is calculated at z = L at the power delivery port. $R_1$ and $R_2$ are the distributed Bragg reflectors at $z = 0$ and $z = L$.}
\label{Fig:laser}
\end{figure}

The basic schematic of the laser that we study in this paper is shown in Fig.~\ref{Fig:laser}. The laser system includes an active region (Yb-doped fiber) with a length of $L$ and a Bragg reflector ($R_1$) at the pumping port ($Z=0$) and another Bragg reflector ($R_2$) at the signal port ($Z=L$). The rate equations for the Yb ions are based on a quasi-three-level system~\cite{Pask}. We assume that the pump power is strong enough to saturate the gain, and the signal is sufficiently strong to suppress the spontaneous emissions. Both of these assumptions are quite reasonable in high-power fiber laser designs.
%%%%%%%%%%%%%%%
%%%%%%%%%%%%%%%

For continuous wave (CW) lasers, the upper manifold populations is given by~\cite{Kelson}
\begin{align}
\label{n2}
\dfrac{N_2(z)}{N}=\dfrac{\dfrac{\Gamma_p \sigma^a_p \widetilde{P_p}(z)}{h \nu_p A}+\dfrac{\Gamma_s \sigma^a_s \widetilde{P_{s~}}(z)}{h \nu_s A}}{\dfrac{\Gamma_p \sigma^{ae}_p \widetilde{P_p}(z)}{h \nu_p A}+\dfrac{1}{\tau}+\dfrac{\Gamma_s \sigma^{ae}_s \widetilde{P_{s~}}(z)}{h \nu_s A}},
\end{align}
where for simplicity of expression we have used the following definitions:
\begin{align}
\widetilde{P_p}(z)\,:=\,P_p^+(z)+P_p^-(z),\quad &&\sigma^{ae}_p\,:=\,\sigma^{a}_p + \sigma^{e}_p ,\\
\widetilde{P_{s~}}(z)\,:=\,P_s^+(z)+P_s^-(z),\quad &&\sigma^{ae}_s\,:=\,\sigma^{a}_s + \sigma^{e}_s.
\end{align}
Equation~\ref{n2} describes the variation of the upper-level population density of the Yb$^{+3}$ ions along the fiber through its dependance on
the $z$-varying pump and signal powers.
The differential equations for forward pump propagation, $P_p^+(z)$, and backward pump propagation, $P_p^-(z)$, are given by
\begin{align}
\label{pumpp}
+\dfrac{dP_p^+(z)}{dz}= \Gamma_p (\sigma^{ae}_p N_2(z)-\sigma^a_p N) P_p^+(z) - \alpha_p P_p^+(z),\\
\label{pumpn}
-\dfrac{dP_p^-(z)}{dz}= \Gamma_p (\sigma^{ae}_p N_2(z)-\sigma^a_p N) P_p^-(z) - \alpha_p P_p^-(z).
\end{align}
The rate equations for forward signal propagation, $P_s^+(z)$, and backward signal propagation, $P_s^-(z)$, are also given by
\begin{align}
\label{sigp}
+\dfrac{dP_s^+(z)}{dz}= \Gamma_s (\sigma^{ae}_s N_2(z)-\sigma^a_s N) P_s^+(z) - \alpha_s P_s^+(z),\\
\label{sign}
-\dfrac{dP_s^-(z)}{dz}= \Gamma_s (\sigma^{ae}_s N_2(z)-\sigma^a_s N) P_s^-(z) - \alpha_s P_s^-(z).
\end{align}
$N$ is the total Yb$^{+3}$ concentration, which is assumed to be constant along the fiber laser, $\nu_s$ ($\nu_p$) is the signal (pump) frequency, 
$\sigma^{a}_s$ ($\sigma^{a}_p$) is the absorption cross section at the signal (pump) wavelength, $\sigma^{e}_s$ ($\sigma^{e}_p$) is the emission cross section at signal (pump)
wavelength, $\tau$ is the upper manifold lifetime, $A$ is the cross-sectional area of the optical fiber core, and $h$ is the Planck constant. 
The fraction of pump power that is coupled to the doped core of the gain fiber is represented by $\Gamma_p$. In a double-cladding configuration, where the pump mode is fully 
scrambled, $\Gamma_p$ can be approximated as the ratio of the doped core area 
to the area of the inner cladding. $\Gamma_s$ is the fraction of the signal power that overlaps the doped core area. Because the core of the optical fiber is single-mode, 
the power filling factor can be easily approximated using an analytical formula presented in Ref.~\cite{agrawal2012fiber} based on the nearly Gaussian profile of the ${\rm LP}_{01}$ mode
in a step-index fiber.
%%%%%%%%%%%%%%%%%%%%%%%%%%%%%%
\section{Analytical solution}
%%%%%%%%%%%%%%%%%%%%%%%%%%%%%%
As already highlighted, the proposed solution applied to the high power laser regime, where the signal power circulating in the cavity is much larger 
than the saturation power of the fiber, i.e. $\widetilde{P_{s~}}\gg P^{\rm sat}_{s}$ and $P^{\rm sat}_{s}=Ah\nu_s/\sigma^{ae}_s\tau$. Because
the total signal power is large enough to saturate the gain at each $z$ location, from Eq.~\ref{n2} combined with the fact that $\sigma^{e}_s\gg\sigma^{a}_s$,
we have $N_2(z) \ll N$~\cite{Kelson, Kelson:99}. For simplicity, we assume that the pump reflection from the second mirror is negligible; however,
the extension to the more general case can be readily implemented. Taking these assumptions into account, Eq.~\ref{pumpp} can be written as
\begin{align}\label{pumppos}
+\dfrac{dP_p^+(z)}{dz}\approx -(\Gamma_p \sigma^a_p N + \alpha_p) P_p^+(z),
\end{align}
The solution to Eq.~\ref{pumppos} can be expressed as 
\begin{align}\label{ppumpprop}
P_p^+(z)=P_p^+(0) e^{- \alpha z},
\end{align}
where we have defined $\alpha$ according to
\begin{align}
\alpha\,:=\,(\Gamma_p \sigma^a_p N + \alpha_p).
\end{align}

We will use Eq.~\ref{pumppos} to account for the propagation of the pump power. We next 
consider the derivation of the relevant equation for the propagation of the forward- and
backward-moving signal power. Equations~\ref{sigp} through \ref{sign} are subject to the following boundary conditions at the location of the mirrors
\begin{align}\label{bound1}
P_s^+(0)= R_1 P_s^-(0),\\
P_s^-(L)= R_2 P_s^+(L).
\end{align}
Using Eq.~\ref{sigp} and, \ref{sign}, we can readily show that
\begin{align}
\label{pzpz1}
\dfrac{dP_s^+(z)}{P_s^+(z)}+\dfrac{dP_s^-(z)}{P_s^-(z)}=0,
\end{align}
which can be used to show that the $z$-derivative of $P_s^+(z) P_s^-(z)$ vanishes, therefore, it is constant along the fiber. In other words,
\begin{align}
\label{pzpz}
P_s^+(0) P_s^-(0)=P_s^+(L) P_s^-(L)=R_1 {P_s^-(0)}^2.
\end{align}

In Eq.~\ref{n2}, when we multiply the sides of the fractions, one of the terms is in the form of $N_2(z)/\tau$; in the Appendix, we have shown that
the variation of $N_2$ with $z$ is very slow and $N_2(z)/\tau$ can be reliably replaced by $\overline{N_2(z)}/\tau$, where $\overline{N_2(z)}$ is
the $z$-averaged value of $N_2(z)$. In Eq.~\ref{n2}, if we replace the $N_2(z)/\tau$ with $\overline{N_2(z)}/\tau$, while keeping the $z$-dependence of other
$N_2(z)$ terms, we arrive at
\begin{align}\label{rewrite}
&\frac{\tau}{h \nu_p A} \big(\Gamma_p (\sigma^{ae}_p N_2(z) - \sigma^a_p N)\big) \widetilde{P_p}+\\\nonumber
&\frac{\tau}{h \nu_s A} \big(\Gamma_s (\sigma^{ae}_s N_2(z) - \sigma^a_s N)\big) \widetilde{P_{s~}} +\overline{N_2}=0.
\end{align}
Using the pump and signal propagation equations directly, we obtain
\begin{align}\label{difp}
(\frac{dP_p^+}{dz}-\frac{dP_p^-}{dz})+ \alpha_p \widetilde{P_p}&= \Gamma_p (\sigma^{ae}_p N_2(z) - \sigma^a_p N) \widetilde{P_p}~,\\
(\frac{dP_s^+}{dz}-\frac{dP_s^-}{dz})+ \alpha_s \widetilde{P_{s~}}&= \Gamma_s (\sigma^{ae}_s N_2(z) - \sigma^a_s N) \widetilde{P_{s~}}\label{difs}.
\end{align}
Inserting Eq.~\ref{difp}, and Eq.~\ref{difs} into Eq.~\ref{rewrite} results in
\begin{align}\label{difp2}
&\frac{\tau}{h \nu_s A} \Big((\frac{dP_s^+}{dz}-\frac{dP_s^-}{dz})+ \alpha_s \widetilde{P_{s~}}\Big)+ \\\nonumber
&\frac{\tau}{h \nu_p A} \Big((\frac{dP_p^+}{dz}-\frac{dP_p^-}{dz})+ \alpha_p \widetilde{P_p}\Big)+ \overline{N_2}=0.
\end{align}
%%%%%%%%%%%%%%%
Using Eq.~\ref{ppumpprop} and the fact that $P_p^-(z)=0$, Eq.~\ref{difp2} can be simplified as
\begin{align}\label{ana}
\nonumber
&\big(\frac{dP_s^+}{dz}-\frac{dP_s^-}{dz}\big)+ \alpha_s \big(P_s^+ + P_s^-\big) + \frac{\nu_s}{\nu_p} e^{- \alpha z} P_p^+(0) \big(\alpha_p-\alpha\big)\\
&+\frac{h \nu_s A \overline{N_2}}{\tau}=0.
\end{align}
%%%%%%%%%%%%%%%%
In order to solve Eq.~\ref{ana}, we first assume that $\alpha_s \approx 0$ and then include the loss term as a perturbation. 
This is permissible because the optical attenuation in optical fibers is very small. In the absence of $\alpha_s$, we can integrate Eq.~\ref{ana} and obtain
\begin{align}\label{mana2}
&P_s^+(z)-P_s^+(0)-P_s^-(z)+P_s^-(0)\\\nonumber
&+\frac{\nu_s}{\nu_p} P_p^+(0)(\alpha_p-\alpha)(\frac{e^{- \alpha z} -1}{- \alpha}) + \frac{h \nu_s A \overline{N_2} z}{\tau}=0.
\end{align}
%%%%%%%%%%%%%%%
Using the boundary conditions (Eq.~\ref{bound1}) and implementing Eq.~\ref{pzpz}, Eq.~\ref{mana2} can be written in following form:
\begin{align}\label{mainquad}
{P_s^+(z)}^2 &+ \Big((1-R_1)P_s^-(0) - \frac{\nu_s}{\nu_p} P_p^+(0) (1-\frac{\alpha_p}{\alpha}) (1-e^{-\alpha z})\\\nonumber
&+ \frac{h \nu_s A \overline{N_2} z}{\tau}\Big) P_s^+(z) - R_1 {P_s^-(0)}^2=0.
\end{align}
Equation~\ref{mainquad} is a quadratic polynomial equation in $P_s^+(z)$ and can be readily solved for $P_s^+(z)$. We can write this equation formally as
\begin{align}
\label{qq1}
X^2 + 2 b X -c =0,\qquad X=P_s^+(z),
\end{align}
where 
\begin{align}\label{quadratic}
&b:= \frac{(1-R_1)P_s^-(0)}{2} - \frac{\nu_s}{2\nu_p} P_p^+(0) (1-\frac{\alpha_p}{\alpha}) (1-e^{-\alpha z})\\ 
\nonumber
&\quad+\frac{h \nu_s A \overline{N_2} z}{2 \tau},\\\nonumber
&c:= R_1 {P_s^-(0)}^2.
\end{align}
The relevant solution to the quadratic polynomial equation that will be on interest to the problem is expressed as
\begin{align}
\label{solX}
X\equiv P_s^+(z) =-b + \sqrt{b^2 + c}.
\end{align}
On the other hand, using $P_s^+(z) P_s^-(z)= R_1 {P_s^-(0)}^2$, we can show that $P_s^+(z) P_s^-(z)= c$. If we define $X'\equiv P_s^-(z)$, we then have $X^\prime=c/X$, which results in
the following quadratic polynomila equation for $X^\prime$:
\begin{align}\label{qq2}
X'^2 - 2 b X' -c =0.
\end{align}
The relevant solution of Eq.~\ref{qq2} is given by 
\begin{align}
\label{solXp}
X'\equiv P_s^-(z) =b + \sqrt{b^2 + c}.
\end{align}
Equations~\ref{solX} and~\ref{solXp} provide the forward- and backward-propagating signal powers; these solutions in combination with Eq.~\ref{n2} and Eq.~\ref{ppumpprop}
present the complete solution to the laser problem. However, we still need to find the value of $P_s^-(0)$ that shows up in $b$ parameter of Eq.~\ref{quadratic} based on the
design parameters of the laser, which is what we will do next.

As explained before, we would now like to find $P_s^-(0)$.
Let's put the two solutions that we found together
\begin{align}
\begin{cases}\label{twoeq}
P_s^+(z) =-b + \sqrt{b^2 + R_1 {P_s^-(0)}^2},\\
P_s^-(z) =+b + \sqrt{b^2 + R_1 {P_s^-(0)}^2}.
\end{cases}
\end{align}
We would like to evaluate these equations at $z=L$, noting that $b$ is also a function of $z$, so we use $b_L=b(z=L)$ in these equations.
If we apply the following boundary conditions
\begin{align}\label{boun}
\begin{cases}
P_s^+(L) P_s^-(L) =P_s^+(0) P_s^-(0),\\
P_s^-(L)=R_2 P_s^+(L),\\
P_s^+(0)=R_1 P_s^-(0),
\end{cases}
\end{align}
which result in
\begin{align}\label{boun}
\begin{cases}
R_2 {P_s^+(L)}^2 = R_1 {P_s^-(0)}^2,\\
P_s^+(L)=\sqrt{R_1R_2^{-1}} P_s^-(0),
\end{cases}
\end{align}
then we obtain the following quadratic polynomial equation for $P_s^+(L)$:
\begin{align}\label{quadps}
{P_s^+(L)}^2 + 2 b_L P_s^+(L) - c =0.
\end{align}
For simplicity the part of $b(L)$ that contains $P_s^-(0)$ can be separated by the following definition:
\begin{align}
&b(L)\equiv \frac{(1-R_1)P_s^-(0)}{2}+ f(L),\\
&f(L)\equiv- \frac{\nu_s}{2\nu_p} P_p^+(0) (1-\frac{\alpha_p}{\alpha}) (1-e^{-\alpha L}) + \frac{h \nu_s A \overline{N_2} L}{2 \tau}.
\end{align}
By inserting $P_s^+(L)=\sqrt{R_1/R_2} P_s^-(0)$ into Eq.~\ref{quadps} the following equation can be obtained:
\begin{align}\label{psl}
&\dfrac{R_1}{R_2} {P_s^-(0)}^2 + 2 \left(\dfrac{1-R_1}{2} P_s^-(0) + f(L)\right) \sqrt{\dfrac{R_1}{R_2}}P_s^-(0)\\\nonumber
&-R_1 {P_s^+(0)}^2=0.
\end{align}
Dividing both side of the Eq.~\ref{psl} by $P_s^-(0)$ results in:
\begin{align}\nonumber
&(\dfrac{R_1}{R_2} + (1-R_1) \sqrt{\dfrac{R_1}{R_2}} - R_1) P_s^-(0) + 2 f(L) \sqrt{R_1}{R_2}=0,\\\nonumber
&\Rightarrow (\sqrt{R_1 R_2} + (R_1-1) - \sqrt{\dfrac{R_1}{R_2}})P_s^-(0) = 2 f(L),\\\nonumber
&\Rightarrow (\sqrt{R1} R_2 + \sqrt{R_2} R_1 - \sqrt{R1}-\sqrt{R2}) P_s^-(0)= 2 \sqrt{R_2} f(L).\\
\end{align}
%%%%%%%%%%%%%%%%%%%
Finally $P_s^-(0)$ can be expressed as:
%%%%%%%%%%%%%%
\begin{align}
\label{ps00app}
&P_s^-(0)= \frac{\sqrt{R_2}}{\left[\sqrt{R_1} (1-R_2) + \sqrt{R_2}(1-R_1) \right]}\\\nonumber
&\times \left[(\frac{\nu_s}{\nu_p})P_p^+(0) (1-\frac{\alpha_p}{\alpha}) (1-e^{-\alpha L}) -  \frac{h \nu_s A \overline{N_2} L}{\tau} \right].
\end{align}
%%%%%%%%%%%%%%
In the Appendix, Eq.~\ref{navg}, we have an expression for $\overline{N_2} L$ that can be substituted in Eq.~\ref{ps00app} to obtain
%%%%%%%%%%%%%%
\begin{align}
\label{ps0app}
&P_s^-(0)=\frac{\sqrt{R_2}}{\left[\sqrt{R_1} (1-R_2) + \sqrt{R_2}(1-R_1) \right]}\\\nonumber
&\times \Bigg[ (\frac{\nu_s}{\nu_p})P_p^+(0) (1-\frac{\alpha_p}{\alpha}) (1-e^{-\alpha L}) \\\nonumber 
&- \frac{h \nu_s A}{\tau} (\frac{\ln \frac{1}{\sqrt{R_1 R_2}} +(\Gamma_s \sigma^a_s N + \alpha_s ) L}{\Gamma_s (\sigma^e_s + \sigma^a_s)}) \Bigg].
\end{align}
%%%%%%%%%%%%%%
All the parameters used in Eq.~\ref{ps0app} are the known laser parameters.
Equation~\ref{ps0app} shows how $P_s^-(0)$ depends on the laser parameters. It is interesting to note that even without considering the complete solution of the 
signal propagation through the laser, one can see how the value of $P_s^-(0)$ depends the mirror reflectivities $R_1$ and $R_2$, or how the length of the fiber, $L$, affects
the value of $P_s^-(0)$. By plugging the value of $P_s^-(0)$ from Eq.~\ref{ps0app} in Eq.~\ref{twoeq}, we obtain the full $z$-dependence of $P_s^+$ and $P_s^-$. However,
for the optimization of a laser's performance using the output signal, we can just use the value of $P_s^-(0)$ in Eqs.~\ref{bound1}, and~\ref{pzpz} and calculate 
$P_s^+(L)$ and $P_s^-(L)$ directly.   

\begin{figure}[H]
\centering
\includegraphics[width=3.3 in]{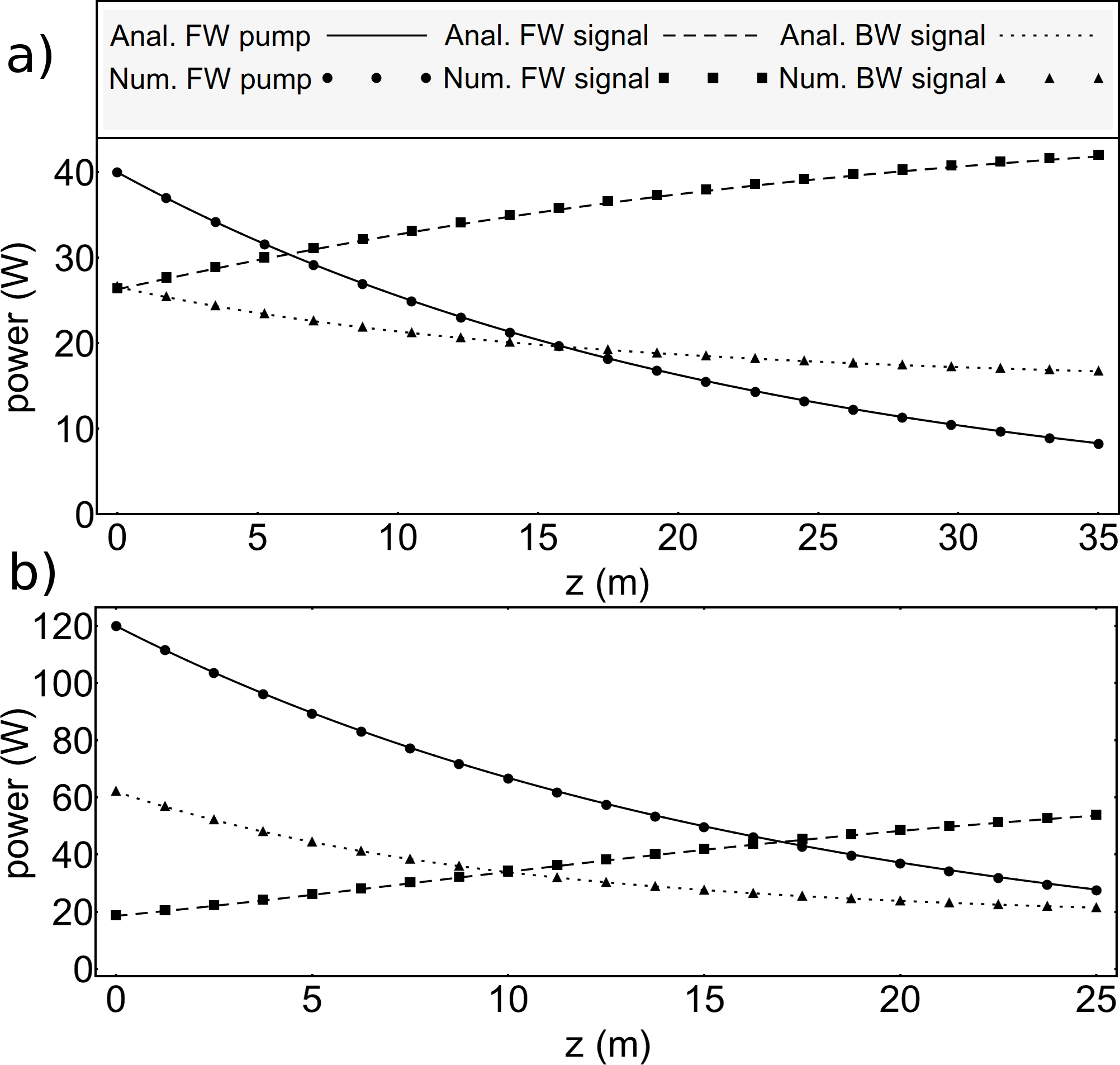}
\caption{a) Comparison of the propagation of the analytical forward pump (Anal. FW pump), analytical forward signal (Anal. FW signal), and analytical backward signal (Anal. BW signal) with their exact numerical counterpart solutions which are the exact numerical forward pump (Num. FW pump), exact numerical forward signal (Num. FW signal), and exact numerical backward signal (Num. BW signal) for the set of Value 1 which are represented in Table~\ref{tab:values}. b) A similar graph for the set of Value 2.}
\label{Fig:anaexac}
\end{figure}

We can now compare our analytical solutions for $P_p^+(z)$, $P_s^+(z)$, and $P_s^-(z)$ with direct numerical simulations. These constitute the full characterization of the laser.
The values of the parameters that are used for the laser system are given in Table.~\ref{tab:values}.
%%%%%%%%%%%%%%
\begin{table}[H]
\centering
\caption{\bf YDCFL parameters}
\scalebox{0.7}{
\begin{tabular}{cccc}
 \hline
 Symbol & Parameter & Value 1 & Value 2 \\
 \hline
$A$ & Core area & $5.0 \times 10^{-11} m^2$ & $5.0 \times 10^{-11} m^2$\\
$\Gamma_s$ & Signal power filling factor & 0.82 & 0.82\\
$\Gamma_p$ & Pump power filling factor & $1.2 \times 10^{-3}$ & $1.2 \times 10^{-3}$\\
$N_0$ & $Yb^{+3}$ concentration & $6.0 \times 10^{25} m^{-3}$& $6.0 \times 10^{25} m^{-3}$\\
$\tau$ & Radiative lifetime & 1.0 ms & 1.0 ms\\

$\sigma^a_p$ & Pump absorption cross section & $6.0 \times 10^{-25} m^{-2}$ & $6.0 \times 10^{-25} m^{-2}$\\
$\sigma^e_p$ & Pump emission cross section & $2.5 \times 10^{-26} m^{-2}$ & $2.5 \times 10^{-26} m^{-2}$\\
$\sigma^a_s$ & Signal absorption cross section & $1.4 \times 10^{-27} m^{-2}$ & $1.4 \times 10^{-27} m^{-2}$\\
$\sigma^e_s$ & Signal emission cross section & $2.0 \times 10^{-25} m^{-2}$ & $2.0 \times 10^{-25} m^{-2}$\\

$\alpha_p$ & Background absorption & $6 \times 10^{-3} m^{-1}$ & $3 \times 10^{-3} m^{-1}$\\
$\lambda_p$ & Pump wavelength & $920~nm$  & $920~nm$  \\
$\lambda_s$ & Signal wavelength & $1090~nm$  & $1090~nm$  \\
$R_1$ & First reflector & $99 \%$ & $30 \%$\\
$R_2$ & Second reflector & $4 \%$ & $40 \%$\\
$L$ & Fiber length & $35 m$ & $60 m$\\
$P_p^+(0)$ & Pump power & $40 W$ & $120 W$\\

\end{tabular}}
  \label{tab:values}
\end{table} 
%%%%%%%%%%%%%%
Figure~\ref{Fig:anaexac} shows a detailed comparison of the analytical versus numerical solution and the agreement is excellent.
Figure~\ref{Fig:anaexac}a corresponds to the set of parameters labeled as Value 1 in Table~\ref{tab:values}, while Fig.~\ref{Fig:anaexac}b 
corresponds to Value 2.
%%%%%%%%%%%%%%
%%%%%%%%%%%%%%

\begin{figure}[H]
\centering
\includegraphics[width=3.3 in]{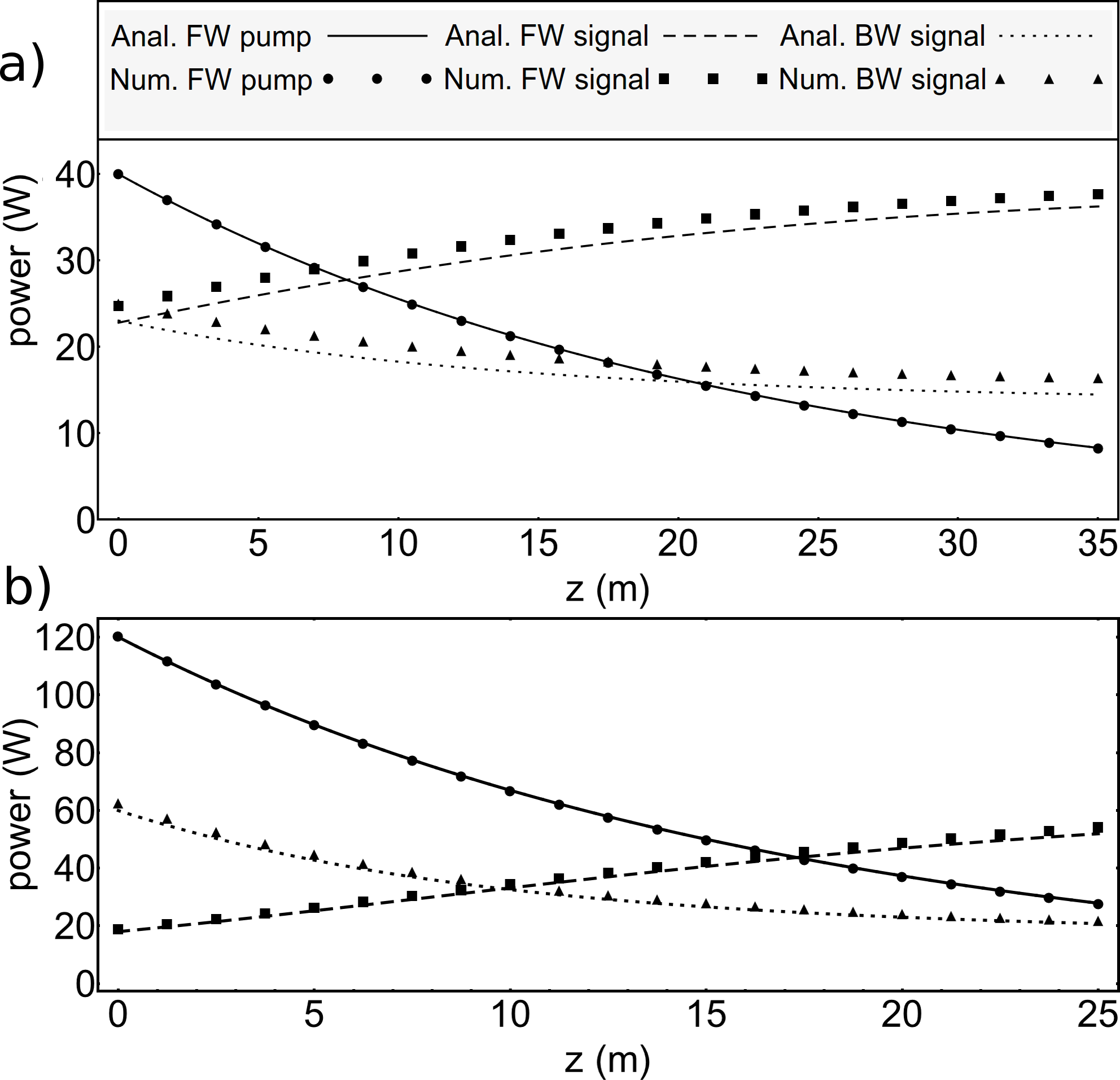}
\caption{a) Results of the analytical solution of a lossy fiber lasers which is a comparison of the propagation of the analytical forward pump (Anal. FW pump), analytical forward signal (Anal. FW signal), and analytical backward signal (Anal. BW signal) with their exact numerical counterpart solutions which are the exact numerical forward pump (Num. FW pump), exact numerical forward signal (Num. FW signal), and exact numerical backward signal (Num. BW signal) for the set of Value 1 which are represented in Table~\ref{tab:values}. b) A similar graph for the set of Value 2.}
\label{Fig:anaexacloss}
\end{figure}

As it is shown in Fig.~\ref{Fig:anaexac}, the analytical solution and the exact numerical solution are in very good agreement; however, the model assumes that the parasitic attenuation of 
the signal $\alpha_s$ is negligible. While in most laser systems this assumption can be acceptable, here we extend the analytical formalism to include $\alpha_s$ using
first-order perturbation theory. If we include the $\alpha_s$ term, we can write (see Eq.~\ref{ana}):
\begin{align}\nonumber
&P_s^+(z)-P_s^+(0)-P_s^-(z)+P_s^-(0)+ \frac{\nu_s}{\nu_p} P_p^+(0)(\alpha_p-\alpha)\Big(\dfrac{e^{- \alpha z} -1}{- \alpha}\Big) \\
&+\frac{h \nu_s A \overline{N_2} z}{\tau}+ \alpha_s \int_{0}^{z} (P_s^+(z)+P_s^-(z)) dz =0.
\end{align}
We next define $\delta$ as
\begin{align}
\label{Eq:deltadef}
\delta = (\dfrac{\alpha_s}{2}) \int_{0}^{z} (P_s^+(z)+P_s^-(z)) dz,
\end{align}
which is a small perturbation to the main Eq.~\ref{quadratic}. This perturbation in effect changes the value of the $b$-parameter 
in Eqs.~\ref{qq1} and~\ref{qq2} to $b+\delta$, which modifies Eq.~\ref{twoeq} to the new form of
\begin{align}
\begin{cases}\label{alpha}
P_{s\delta}^+(z) =-(b+\delta) + \sqrt{(b+\delta)^2 + R_1 {P_{s\delta}^-(0)}^2},\\
P_{s\delta}^-(z) =+(b+\delta) + \sqrt{(b+\delta)^2 + R_1 {P_{s\delta}^-(0)}^2}.
\end{cases}
\end{align}
To find an expression for $\delta$ in the first-order perturbation theory, we need to use $P_s^+(z)$ and $P_s^-(z)$ from Eq.~\ref{twoeq} (corresponding to $\alpha_s=0$) in 
Eq.~\ref{Eq:deltadef}. We obtain
\begin{align}\label{del}
\delta=\alpha_s \int_{0}^{z} \sqrt{b(z)^2 + R_1 {P_s^-(0)}^2} dz.
\end{align}
In general, $b$ is a complicated function of $z$ and the integral in Eq.~\ref{del} cannot be simplified analytically.
To proceed further analytically, we can implement the midpoint rule (rectangle rule) for the integration and approximate $\delta$ as
\begin{align}\label{del}
\delta(z)\approx\alpha_s\sqrt{b(z/2)^2 + R_1 {P_s^-(0)}^2}\,z,
\end{align}
where $b$ is evaluated at the midpoint $z/2$. We next insert the value of $b(L)$ from Eq.~\ref{del} in Eq.~\ref{alpha} and obtain 
$P_{s\delta}^+(L)$. Note that we use $P_{s}^-(0)$ under the square root in Eq.~\ref{alpha} to comply with the first order perturbation. 
We then apply the boundary condition $P_{s\delta}^-(0)=\sqrt{R_2/R_1}P_{s\delta}^+(L)$ and calculate the first-order improved 
$P_{s\delta}^-(0)$, which is then used in Eq.~\ref{alpha} to obtain the full $z$-dependence of $P_{s\delta}^+$ and $P_{s\delta}^-$. 
This procedure results in a rather accurate account of the forward- and backward- moving signal propagation.
Figure~\ref{Fig:anaexacloss}a corresponds to the set of parameters labeled as Value 1 in Table~\ref{tab:values}, while Fig.~\ref{Fig:anaexacloss}b 
corresponds to Value 2. The signal attenuation parameter is taken to be equal to that of the pump, i.e. $\alpha_s=\alpha_p$, in either case.  
%%%%%%%%%%%%%%%%%%%%%%%%%%%%%%%%%%%%%%%%%%%%%%%%%%%%%%%%%%%%%%%%%%%%%%%%%%%%%%%%%%%%
%%%%%%%%%%%%%%%%%%%%%%%%%%%%%%%%%%%%%%%%%%%%%%%%%%%%%%%%%%%%%%%%%%%%%%%%%%%%%%%%%%%%
\section{Conclusion}
%%%%%%%%%%%%%%%%%%%%%%%%%%%%%%%%%%%%%%%%%%%%%%%%%%%%%%%%%%%%%%%%%%%%%%%%%%%%%%%%%%%%
%%%%%%%%%%%%%%%%%%%%%%%%%%%%%%%%%%%%%%%%%%%%%%%%%%%%%%%%%%%%%%%%%%%%%%%%%%%%%%%%%%%%
In summary, we have presented a consistent fully analytical solution for the propagation of signal and pump in YDCFL for both lossy and lossless cases. 
The results are in excellent agreement with the direct numerical solution. This study is important because calculating the pump and signal power in a fiber laser 
using a numerical solution involves iterative solutions of coupled differential equations and can be time consuming. An analytical solution is especially
beneficial for multi-parameter optimization in designing a fiber laser. The presence of an analytical solution can significantly speed up optimization algorithms
that targets such metrics as the laser output power, efficiency, heat generation or temperature, or a combination of these parameters. The optimization parameters
may include the choice of the fiber geometry especially the length, fiber dopant concentration, and mirror reflectivities. However, when addressing the
thermal issues such as in a radiation balanced laser (RBL) design~\cite{Bowman,Mobini:18}, pump and signal wavelengths are also included as design parameters, 
making direct numerical optimization a very time-consuming task~\cite{Peysokhan:20}. This study will pave the way to make it possible to find optimally designed lasers over a large design parameter space.
%%%%%%%%%%%%%%%%%%%%%%%
%%%%%%%%%%%%%%%%%%%%%%%
%%%%%%%%%%%%%%%%%%%%%%%
%%%%%%%%%%%%%%%%%%%%%%%
%%%%%%%%%%%%%%%%%%%%%%%
%%%%%%%%%%%%%%%%%%%%%%%
%%%%%%%%%%%%%%%%%%%%%%%
%%%%%%%%%%%%%%%%%%%%%%%
%%%%%%%%%%%%%%%%%%%%%%%
%%%%%%%%%%%%%%%%%%%%%%%
%%%%%%%%%%%%%%%%%%%%%%%
\section{Appendix: Average of $N_2(z)$ over the length}
\label{appendix 1}
\begin{figure}[H]
\centering
\includegraphics[width=3.3 in]{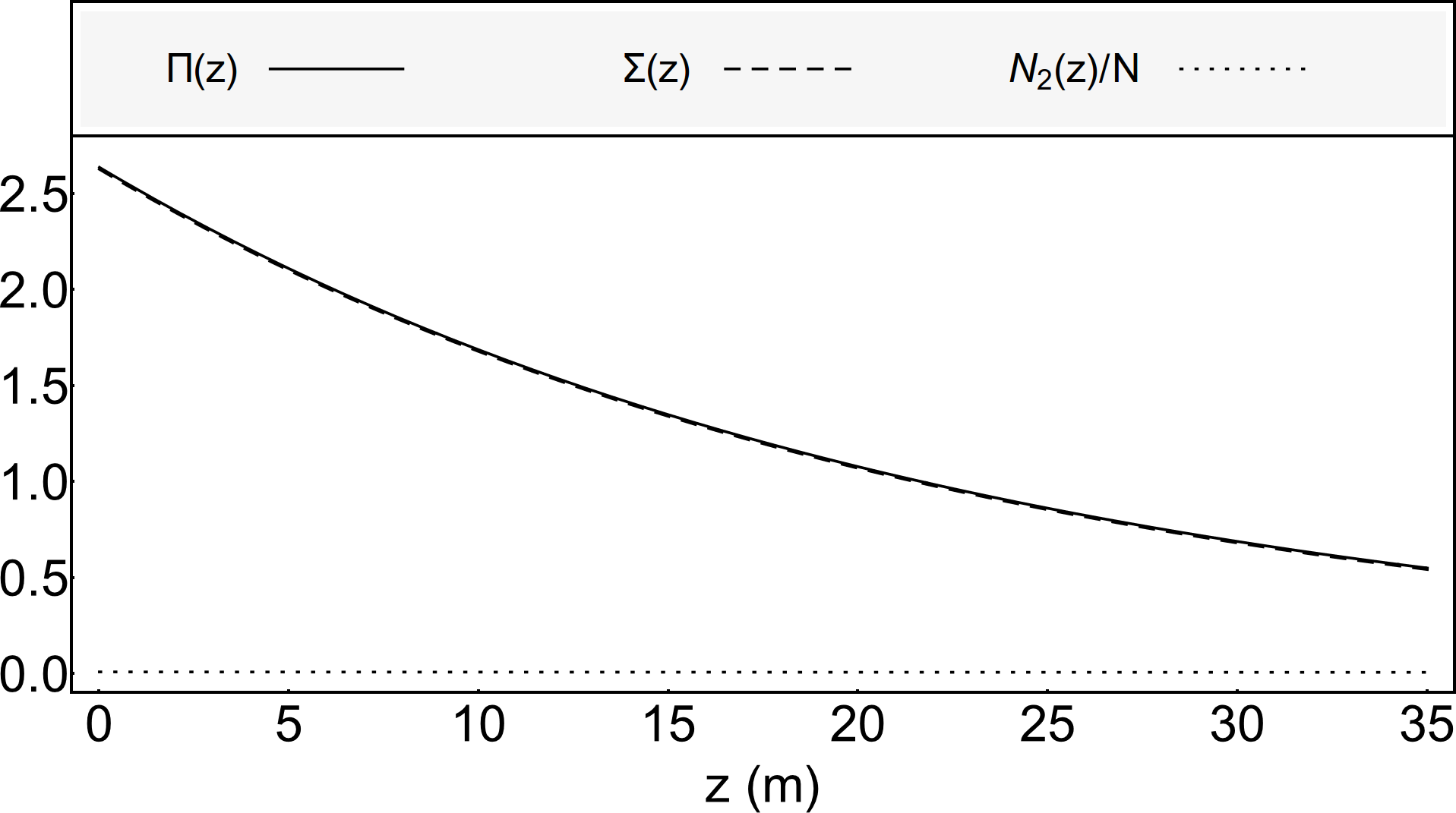}
\caption{Comparison of the $\Pi(z)$, $\Sigma(z)$, and $N_2(z)/N$ along the doped fiber for the Value 1 that are listed in Table~\ref{tab:values}}
\label{Fig:comparison}
\end{figure}

Equation~\ref{n2} can be written in the form of
%%%%%%%%%%%%%%%%%%%%%%%
\begin{align}\label{n2/n}
&\widetilde{P_p}(z)(\frac{\tau \Gamma_p}{h \nu_p A}) \Big(\sigma^a_p-\big(N_2(z)/N\big) \sigma^{ae}_p\Big)\\\nonumber
&=\frac{N_2(z)}{N} + \widetilde{P_{s~}}(\frac{\tau \Gamma_s}{h \nu_s A}) \Big(\big(N_2(z)/N\big) \sigma^{ae}_s-\sigma^a_s\Big),
\end{align}
%%%%%%%%%%%%%%%%%%%%%%%
We also define $\Pi(z)$ and $\Sigma(z)$ as
%%%%%%%%%%%%%%%%%%%%%%%
\begin{align}
\Pi(z)\,:=\,\widetilde{P_p}(z)(\frac{\tau \Gamma_p}{h \nu_p A}) \Big(\sigma^a_p-\big(N_2(z)/N\big) \sigma^{ae}_p\Big),\\ \nonumber
\Sigma(z)\,:=\,\widetilde{P_{s~}}(z)(\frac{\tau \Gamma_s}{h \nu_s A}) \Big(\big(N_2(z)/N\big) \sigma^{ae}_s-\sigma^a_s\Big).
\end{align}
%%%%%%%%%%%%%%%%%%%%%%%
Therefore, we can write Eq.~\ref{n2/n} as $\Pi(z)-\Sigma(z)=N_2(z)/N$. As shown in Fig.~\ref{Fig:comparison}, while
$\Pi(z)$ and $\Sigma(z)$ vary considerably along the $z$-coordinate, the variation of $N_2(z)/N$ along the length of the fiber is very small.
Therefore, it is reasonable to use, instead of $N_2(z)/N$, $\overline{N_2}/N$ which is its average value over the length of the fiber. 
It should be noted that this substitution is just applied to the first term of right-hand side of Eq.~\ref{n2/n} and not to the $N_2(z)/N$ 
term present in $\Pi(z)$ and $\Sigma(z)$.
%%%%%%%%%%%%%%%%%%%%%%% 
%%%%%%%%%%%%%%%%%%%%%%%

The average value of $N_2(z)$ over the fiber length can be expressed as 
%%%%%%%%%%%%%%%%%%%%%%%
\begin{align}
\overline{N_2}= \frac{1}{L} \int_0^L N_2(z) dz.
\end{align}
%%%%%%%%%%%%%%%%%%%%%%%
The best way to represent $\overline{N_2}$ is by utilizing the total gain of the fiber laser system. The total gain of the fiber laser for each round trip is given by
%%%%%%%%%%%%%%%%%%%%%%%
\begin{align}\label{n2a}\nonumber
&G=\int_0^L \frac{dP_s^+(z)}{P_s^+(z)} + \int_L^0 \frac{dP_s^-(z)}{P_s^-(z)}\\\nonumber
&=\int_0^L \Big(\Gamma_s \big(\sigma^{ae}_s N_2(z)-\sigma^a_s N\big) - \alpha_s\Big)dz \\\nonumber
&- \int_L^0 \Big(\Gamma_s \big(\sigma^{ae}_s N_2(z)-\sigma^a_s N\big) - \alpha_s\Big)dz\\\nonumber
&=2 \Gamma_s \sigma^{ae}_s \int_0^L N_2(z) dz - 2 (\Gamma_s \sigma^a_s N + \alpha)L,\\
\end{align}
%%%%%%%%%%%%%%%%%%%%%%%
where $G$ is the total gain of the fiber laser. From the final result of Eq.~\ref{n2a}, the following equation can be derived:
%%%%%%%%%%%%%%%%%%%%%%%
\begin{align}
&\frac{1}{L} \int_0^L N_2(z) dz = \overline{N_2}= \frac{\frac{G}{2 L} + (\Gamma_s \sigma^a_s N + \alpha_s )}{\Gamma_s \sigma^{ae}_s}.
\end{align} 
%%%%%%%%%%%%%%%%%%%%%%%
The gain parameter, $G$ should be evaluated based on laser characters. In each round trip the following relations are
%%%%%%%%%%%%%%%%%%%%%%%
\begin{align}
\label{gain}
R_1 R_2  e^{G}=1 \Rightarrow \frac{G}{2 L} = \frac{1}{L} \ln \frac{1}{\sqrt{R_1 R_2}}.
\end{align}
%%%%%%%%%%%%%%%%%%%%%%%
By combining the Eq.~\ref{n2a} and Eq.~\ref{gain} the following expression can be obtained for the average value of $N_2(z)$:
\begin{align}\label{navg}
\overline{N_2} = \frac{\frac{1}{L} \ln \frac{1}{\sqrt{R_1 R_2}} +(\Gamma_s \sigma^a_s N + \alpha_s )}{\Gamma_s (\sigma^e_s + \sigma^a_s)},
\end{align}
where $\overline{N_2}$ depends only on the known laser parameters.
%%%%%%%%%%%%%%%%%%%%%%%
%%%%%%%%%%%%%%%%%%%%%%%
%%%%%%%%%%%%%%%%%%%%%%%
%%%%%%%%%%%%%%%%%%%%%%%
%%%%%%%%%%%%%%%%%%%%%%%
%%%%%%%%%%%%%%%%%%%%%%%
%%%%%%%%%%%%%%%%%%%%%%%
%%%%%%%%%%%%%%%%%%%%%%%
%%%%%%%%%%%%%%%%%%%%%%%
%%%%%%%%%%%%%%%%%%%%%%%
%%%%%%%%%%%%%%%%%%%%%%%

%%%%%%%%%%%%%%%%%%%%%%%
%%%%%%%%%%%%%%%%%%%%%%%
%%%%%%%%%%%%%%%%%%%%%%%%%%%%%%%%%%%%%%%%%%%%%%%%%%%%%%
%%%%%%%%%%%%%%%%%%%%%%%%%%%%%%%%%%%%%%%%%%%%%%%%%%%%%%
\section*{Acknowledgments}
%%%%%%%%%%%%%%%%%%%%%%%
%%%%%%%%%%%%%%%%%%%%%%%
This material is based upon work supported by the Air Force Office of Scientific Research under award number FA9550-16-1-0362 titled Multidisciplinary Approaches to Radiation Balanced Lasers (MARBLE).

\end{document}